\documentclass[3p,times]{elsarticle}

\usepackage{ecrc}
\usepackage{multirow}
\usepackage[bookmarksnumbered, pdfstartview=FitH,colorlinks,urlcolor=blue, citecolor=blue,linkcolor=blue,] {hyperref}

\volume{00}

\firstpage{1}

\journalname{Nuclear Instruments and Methods in Physics Research Section A}

\runauth{H. Miao, W. Chen}

\jid{procs}

\jnltitlelogo{}

\usepackage{amssymb}
\usepackage{amsmath}
\usepackage{subfigure}

\usepackage[figuresright]{rotating}
\usepackage{setspace}

\begin{document}

\doublespacing


\begin{frontmatter}




\title{Implementation of antiproton generator in Geant4}


	\author[a,b]{~Han Miao\footnote{Email: miaohan@ihep.ac.cn}}
	\author[c]{~Chen Wu}
	\author[a,b]{~Hai-Bo Li}

	\address[a]{\it Institute of High Energy Physics, Beijing 100049, People's Republic of China}
	\address[b]{\it University of Chinese Academy of Sciences, Beijing 100049, People's Republic of China}
	\address[c]{\it Department of Physics, Osaka University, Toyonaka, Osaka 560-0043, Japan}

\begin{abstract}
\doublespacing
	A generator of antiproton ($\bar{p}$) in Geant4 has been developed using a theoretical model with the consideration of near-threshold treatment and used preliminarily in COMET experiment. The precision of this generator has been verified by comparing with experimental results taken in CERN (1970) and ITEP (1994). By implementing this generator into official data analysis framework of COMET, the yield of $\bar{p}$ in COMET experiment can be evaluated properly. Background caused by $\bar{p}$ can also be estimated by this generator in future, which will instruct the design and optimization. Moreover, this generator can also be used in the simulation of other experiments involving near-threshold $\bar{p}$ generation.
\end{abstract}


\end{frontmatter}


\section{Introduction}
\label{chap:introduction}

Simulation of yields of charged particles, especially the yield of $\bar{p}$, has always been a difficulty in proton-nucleon collisions at low energy especially when center-of-mass (c.m.s) energy $\sqrt{s}<10~\rm{GeV}$ since there are many theoretical methods of approximation near the threshold~\cite{Taylor:1975tm, Frankel:1977sj, Tan:1983kgh, Danielewicz:1990cd, Sibirtsev:1997mq, Duperray:2003bd, diMauro:2014zea} but few precise experiment results for validation in the past. While, simulation of $\bar{p}$ production is usually a necessary work for many fields of high energy physics and astrophysics, for example, evaluation of background, design of $\bar{p}$ source and research on cosmic ray or air shower. Many simulation works have been taken in this field~\cite{Lebedev:2001zz, Kappl:2014hha, Donato:2001ms, Jezabek:2021oxg, STAR:2003oii}, but $\sqrt{s}$ of these works are much higher than 10~GeV.

Recently, various experiments have been proposed or designed to search for new physics using the muon beam produced from the collision of protons and nucleons, including COMET (The COherent Muon to Electron Transition)~\cite{COMET:2009qeh, COMET:2018auw}, Mu2e~\cite{Kutschke:2009zz} and ones in future. Most of these experiments will meet this problem during the optimization. In the case of COMET and Mu2e, limited by the accelerator, the energy of proton beam is set to be 8~GeV ($\sqrt{s} \approx 3.86~{\rm GeV}$), which is in the near-threshold region. $\bar{p}$ production is suppressed at this energy but the near-threshold approximation is necessary to evaluate the background of $\bar{p}$. Therefore, we develop a generator of $\bar{p}$ for more precise simulation. Meanwhile, this generator can also be used in the simulation of future experiments relavant to the $\bar{p}$ production in $p$-nucleon collisions. 

The COherent Muon to Electron Transition (COMET) experiment~\cite{COMET:2009qeh, COMET:2018auw} at the Japan Proton Accelerator Research Complex (J-PARC) in Tokai, Japan, aims to search for the neutrinoless coherent transition of a muon to an electron ($\mu-e$ conversion) in the field of a necleus, with its single event sensitivity of $2\times 10^{-17}$, which is more than four orders of magnitude improvement over than current upper limit of $7\times 10^{-13}$ at $90 \%$ C.L. from the SINDRUM~II experiment~\cite{SINDRUMII:2006dvw}. As COMET experiment seeks for the process out of standard model that will be extramely rare, estimation of background level will be of much significance for the design and optimization. Background caused by $\bar{p}$ with low momenta will be dangerous for detectors and may contaminate the target signals. Since $\bar{p}$ doesn't decay and has relatively larger mass, it cannot be suppressed by the pulse structure of beam. Furthermore, electrons with energy close to the target signal may be produced from the decay of $\pi$ mesons, which are produced by the interaction of $\bar{p}$ in the beam and the materials near the detectors. For COMET experiments, much fewer $\bar{p}$ (less than $1\times 10^{-8}$ of all the produced $\bar{p}$) are able to arrive the detector area as only the secondary beam towards backward direction is collected, so it is of great significance to precisely describe the distribution of $\bar{p}$ in the phase space.

In order to evaluate a reasonable acceptance of each kind of particles in detector area, simulation work is performed using Geant4~\cite{GEANT4:2002zbu}. However, yield of $\bar{p}$ when protons with $8~\rm{GeV}$ hitting target will be close to $0$ in former simulation using the official physics lists, because the hadronic processes of Geant4 depend on the interpolation and extrapolation of experiment results without dealing with specific processes~\cite{Geant4:physics}. This generator extended the physics models by utilizing the theoretical model informed in the following chaper.

\section{Method}

In our generator, $\bar{p}$ are produced by the collisions of incident protons and protons in nuleus, namely, $p+p\rightarrow p+p+p+\bar{p}$. By assuming the dynamics of nucleons follow the Fermi Gas model~\cite{Benhar:1994hw}, which has already been implemented inside Geant4, the momenta of protons in nucleus (Fermi momenta) can be got by sampling the momentum distribution of nucleons.

After determining the momenta of both incident and nucleon protons, we need to get the differential production cross section of $\bar{p}$. During the last several decades, cross sections of charged particles produced in proton-nucleon interactions are measured at various energy points~\cite{Capiluppi:1974rt, E-802:1992lwr, British-Scandinavian-MIT:1976wao, Dekkers:1965zz, Boyarinov:1994tp, Allaby:1970jt, Kiselev:2012sj, Amaldi:1975hot, Chiba:1990kh, Anderson:1967zzc}. While, among these experiments, the lowest c.m.s energy is $4.98~\rm{GeV}$, which is still higher than the case of COMET, and the largest scattering angle in laboratory system is $119^{\circ}$, much smaller than COMET case. For the same reason, cross sections used in Geant4 simulation may not be accurate because of lacking experimental support. To solve this problem, we use a theoretical model developed by Tan et al. in 1983~\cite{Tan:1983kgh} considering the machinism of $\bar{p}$ production, of which the precision has been verified by the comparisons of Chapter~3 between the model and experiments at various cases.

In the model mentioned above, the differential cross section is described by the following formula:

\begin{equation}
	\label{equ:diff_cross_section}
	(E\cdot \frac{d^{3}\sigma}{d^{3}p})_{LE}(x_R, x_T, p_t) = (E\cdot\frac{d^{3}\sigma}{d^{3}p})_{RS}(x_R, p_t) \cdot R(x_T, \sqrt{s})
\end{equation}
	
In this formula, $\sqrt{s}$ is the c.m.s energy of $\bar{p}$ and $p_t$ is the transverse momentum of $\bar{p}$. $x_R = E/E_{max}$ refers to the ratio of total energy over the maximum total energy and $x_T=T/T_{max}$ is the ratio of kinetic energy over the maximum kinetic energy in c.m.s. $E_{max}$ and $T_{max}$ are given by the following formula due to the kinematic constrains:

\begin{equation}
        E_{max}= \frac{s-{\bar{M}}_{X}^{2} c^4 +m^2 c^4}{2\sqrt{S}}, ~~{\bar{M}}_{X} = 2.814~\rm{GeV}
\end{equation}

\begin{equation}
        x_T = \frac{x_R-x_m}{1-x_m},~~ x_m = mc^2/E_{max}
\end{equation}

And, $(E\cdot\frac{d^{3}\sigma}{d^{3}p})_{RS}$ refers to the invariant cross section of $p+p \rightarrow h+X$ (h-hadron) when $\sqrt{s} \rightarrow \infty$ called radial-scaling limit according to Feynman's work~\cite{Feynman:1969ej}. In 1976, Taylor et al.~\cite{Taylor:1975tm} found that the radial-scaling limit is already reached at $\sqrt{s} > 10~\rm{GeV}$. In other word, $\bar{p}$ production cross section can be described by $(E\cdot\frac{d^{3}\sigma}{d^{3}p})_{RS}$ above $10~\rm{GeV}$. Below $10~\rm{GeV}$, the cross section in low-energy case $(E\cdot \frac{d^{3}\sigma}{d^{3}p})_{LE}$ can be got by multiplying a correction factor $R(x_T, \sqrt{s})$ to $(E\cdot\frac{d^{3}\sigma}{d^{3}p})_{RS}$.

$(E\cdot\frac{d^{3}\sigma}{d^{3}p})_{RS}$ and the ratio $R(x_T, \sqrt{s})$ can be calculated by the following function according to Ref.~\cite{Tan:1983kgh}:

\begin{equation}
	\label{equa:CS_RS}
	(E\cdot\frac{d^{3}\sigma}{d^{3}p})_{RS} = f(x_R){\rm{exp}}[-(A(x_R)p_t+B(x_R)p_{t}^{2})]~({\rm{mb}}~{\rm{GeV}^{-2}}~c^{3})
\end{equation}

\begin{equation}
	f=a_1 {\rm{exp}}(-a_2 x_R)\theta(a_3-x_R)+(\sigma_{00}-a_1)(1-x_R)^{a_4}
\end{equation}

\begin{equation}
	A = a_1 {\rm{exp}}(-a_6 x_R)+a_7 exp(a_8 x_R)
\end{equation}

\begin{equation}
	B = a_9 {\rm{exp}}[-a_{10} (x_R+a_11)](x_R+a_11)^{a_{12}}
\end{equation}

\begin{equation}
	\theta(u) = \left\{\begin{matrix}
 0 & for~u<0\\
 1 & for~u\ge 0
	\end{matrix}\right.
\end{equation}

\begin{equation}
	R^{-1} = 1-{\rm{exp}}[-(1-{\rm{exp}}(-A(x_T)Q^{B(x_T)}))~{\rm{exp}}(C(x_T)Q-D(x_T))]
\end{equation}

\begin{equation}
	Q=\sqrt{S}-\sqrt{S_t}, ~~\sqrt{S_t} = 3.752~\rm{GeV}
\end{equation}

\begin{equation}
	A=b_1~{\rm{exp}}(-b_2 x_T)
\end{equation}

\begin{equation}
	B=b_3~{\rm{exp}}(b_4x_T)
\end{equation}

\begin{equation}
	C=b_5+b_6 x_T+b_7 x_{T}^{2}
\end{equation}

\begin{equation}
	D=b_8~{\rm{exp}}(b_9 x_T)
\end{equation}
where $a_1,...,a_{12}$ are parameters shown in Table~\ref{tab:parametersa} and $b_1,...,b_9$ are parameters shown in Table~\ref{tab:parametersb}.

\begin{table}[!htbp]
	\doublespacing
	\centering
	\footnotesize
	\begin{tabular}{ccccccccccccc}
		\hline
		\hline
		$\sigma_{00}$ & $a_1$ & $a_2$ & $a_3$ & $a_4$ & $a_5$ & $a_6$ & $a_7$ & $a_8$ & $a_9$ & $a_{10}$ & $a_{11}$ & $a_{12}$ \\
		\hline
		3.15 & $1.05\times10^{-4}$ & 1.01 & 0.5 & 7.90 & 0.465 & $3.70\times10^{-2}$ & 2.31 & $1.40\times10^{-2}$ & $3.02\times10^{-2}$ & 3.19 & 0.399 & 8.39 \\
		\hline
	\end{tabular}
	\label{tab:parametersa}
	\caption{Parameters in the formula of differential cross section from Ref.~\cite{Tan:1983kgh}.}
\end{table}

\begin{table}[!htbp]
	\doublespacing
	\centering
	\footnotesize
	\begin{tabular}{ccccccccc}
		\hline
		\hline
		$b_1$ & $b_2$ & $b_3$ & $b_4$ & $b_5$ & $b_6$ & $b_7$ & $b_8$ & $b_9$ \\
		\hline
		0.306 & 0.120 & 0.0552 & 2.72 & 0.758 & -0.680 & 1.54 & 0.594 & 2.87 \\
		\hline
	\end{tabular}
	\label{tab:parametersb}
	\caption{Parameters in the expression of $R$ from Ref.~\cite{Tan:1983kgh}.}
\end{table}

\begin{figure*}[!htbp]
	\doublespacing
	\centering
	\includegraphics[width=0.50\textwidth]{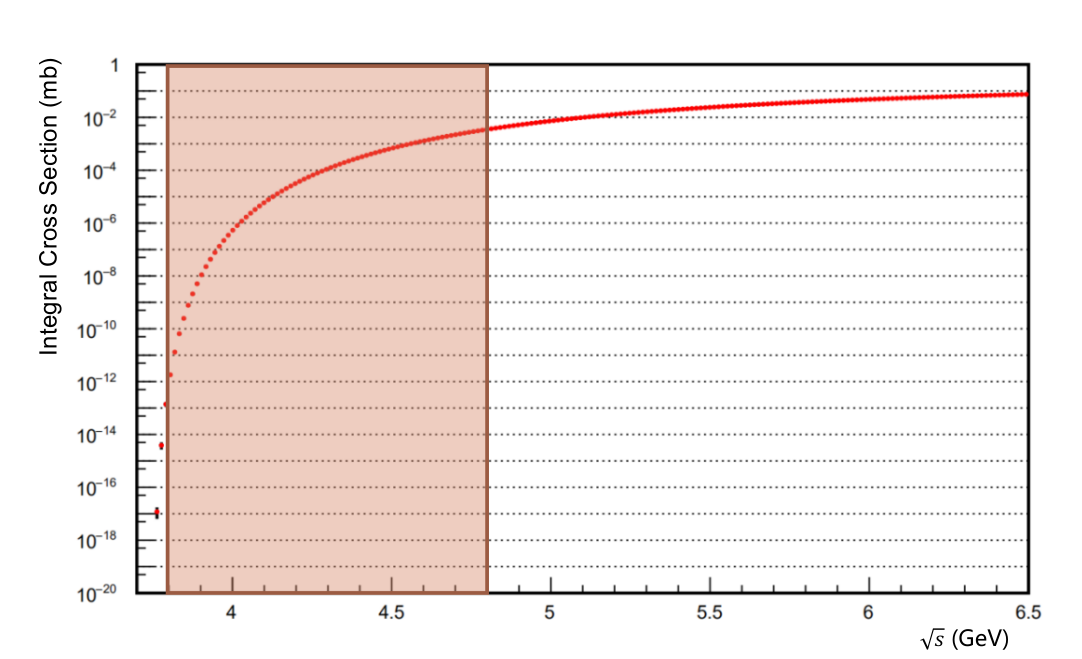}
	\caption{Integral production cross section of $\bar{p}$.}
	\label{fig:total_CS}
\end{figure*}

With the formula shown above, we can get differential cross sections of $\bar{p}$ production. And then integral cross section is available by integrating the differential cross sections within the possible momentum range of $\bar{p}$. The total cross sections within $[3.7, 6.5]~{\rm{GeV}}/c$ have been shown in Figure~\ref{fig:total_CS}, where the red part is the c.m.s. energy in COMET case considering the Fermi momenta. With the properties of material, the mean free path of this process can be calculated, which will be transported to Geant4 kernel to issue a step length.

\begin{figure*}[!htbp]
	\centering
	\subfigure[]{ \includegraphics[width=0.7\textwidth]{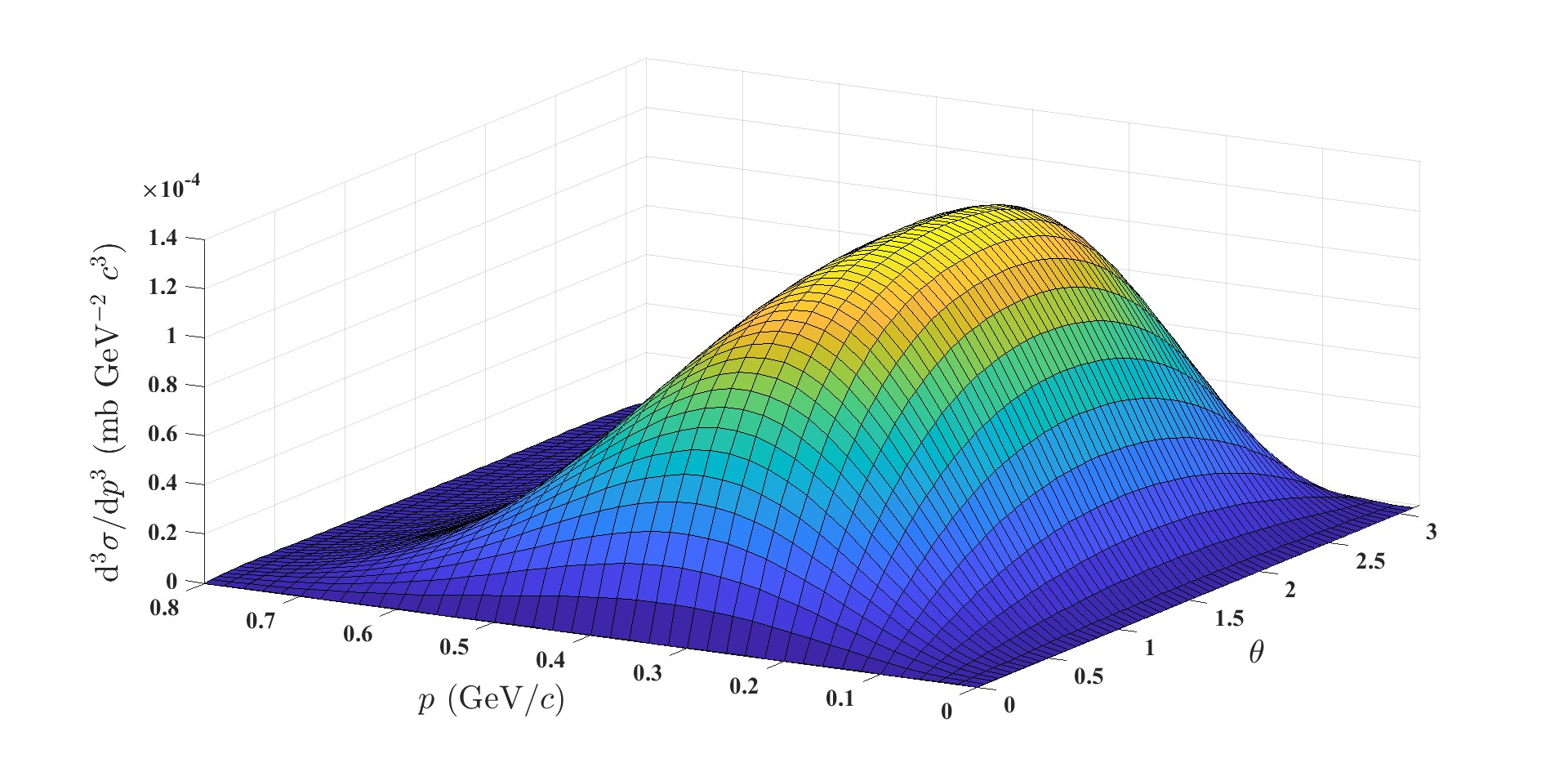}}
	\subfigure[]{ \includegraphics[width=0.7\textwidth]{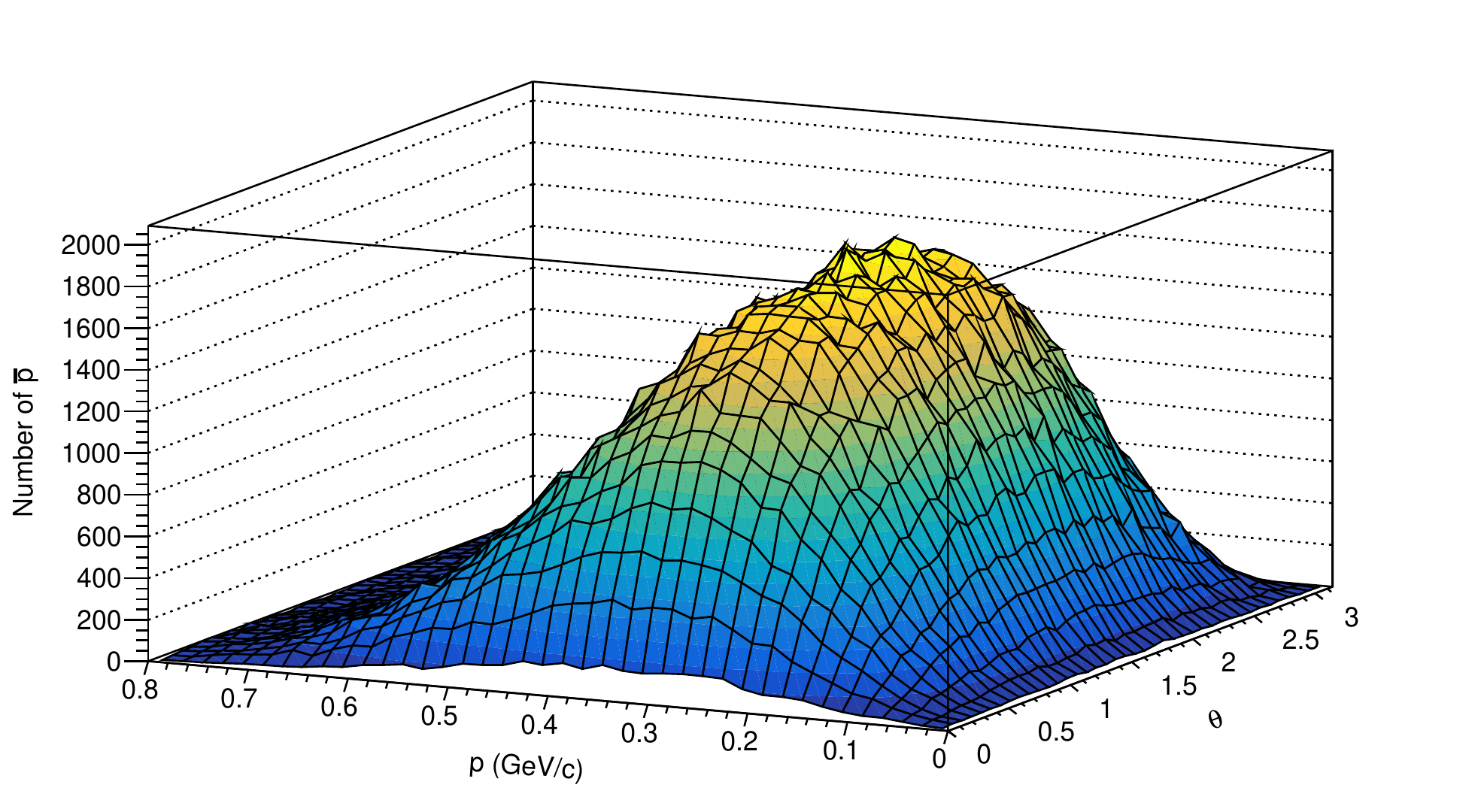}}
	\subfigure[]{ \includegraphics[width=0.7\textwidth]{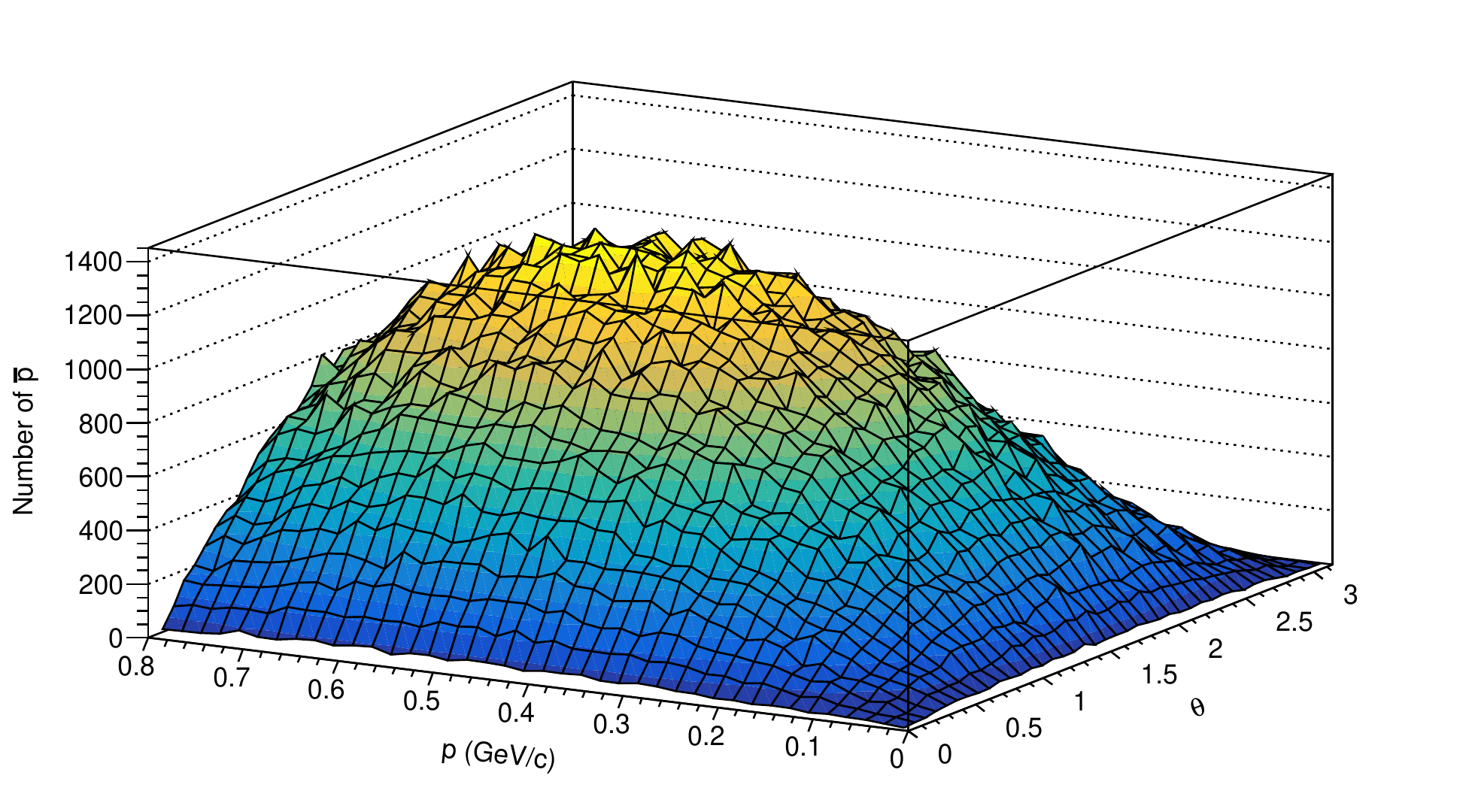}}
	\caption{(a) Theoretical differential cross section when 8 GeV proton hits the nucleon. (b) Momentum distribution got by simulation using the $\bar{p}$ generator in this work ($1\times10^6$ incident proton). (c) Momentum distribution got by sampling phase space ($1\times10^6$ incident proton).}
	\label{fig:momenta_distribution}
\end{figure*}

After this process is determined to occur by Geant4 kernel, the momenta of $\bar{p}$ are got by sampling the differential cross section distribution mentioned above. Figure~\ref{fig:momenta_distribution} (a) shows the theoretical differential cross section and (b) shows the momentum distribution got from simulation. These two distributions are consistent with each other well. For reference, Figure~\ref{fig:momenta_distribution} (c) is the case with the trivial assumption that $\bar{p}$ uniformly distributes in the c.m.s. phase space. Obviously, the trivial assumption is significantly different from the real case.

\section{Comparison with experimental results}
\subsection{Data taken by CERN in 1970}

As mentioned above, several experiments has been performed to measure the production cross sections of charged particles in proton-nucleon interaction~\cite{Capiluppi:1974rt, E-802:1992lwr, British-Scandinavian-MIT:1976wao, Dekkers:1965zz, Boyarinov:1994tp, Allaby:1970jt, Kiselev:2012sj, Amaldi:1975hot, Chiba:1990kh, Anderson:1967zzc}. We firstly pick the experiment of CERN in 1970~\cite{Allaby:1970jt} to compare with our simulation results since the detailed parameters and data of this experiment can be found in a published paper.

\begin{figure}[!htbp]
	\doublespacing
	\centering
	\subfigure[]{ \includegraphics[width=0.45\textwidth]{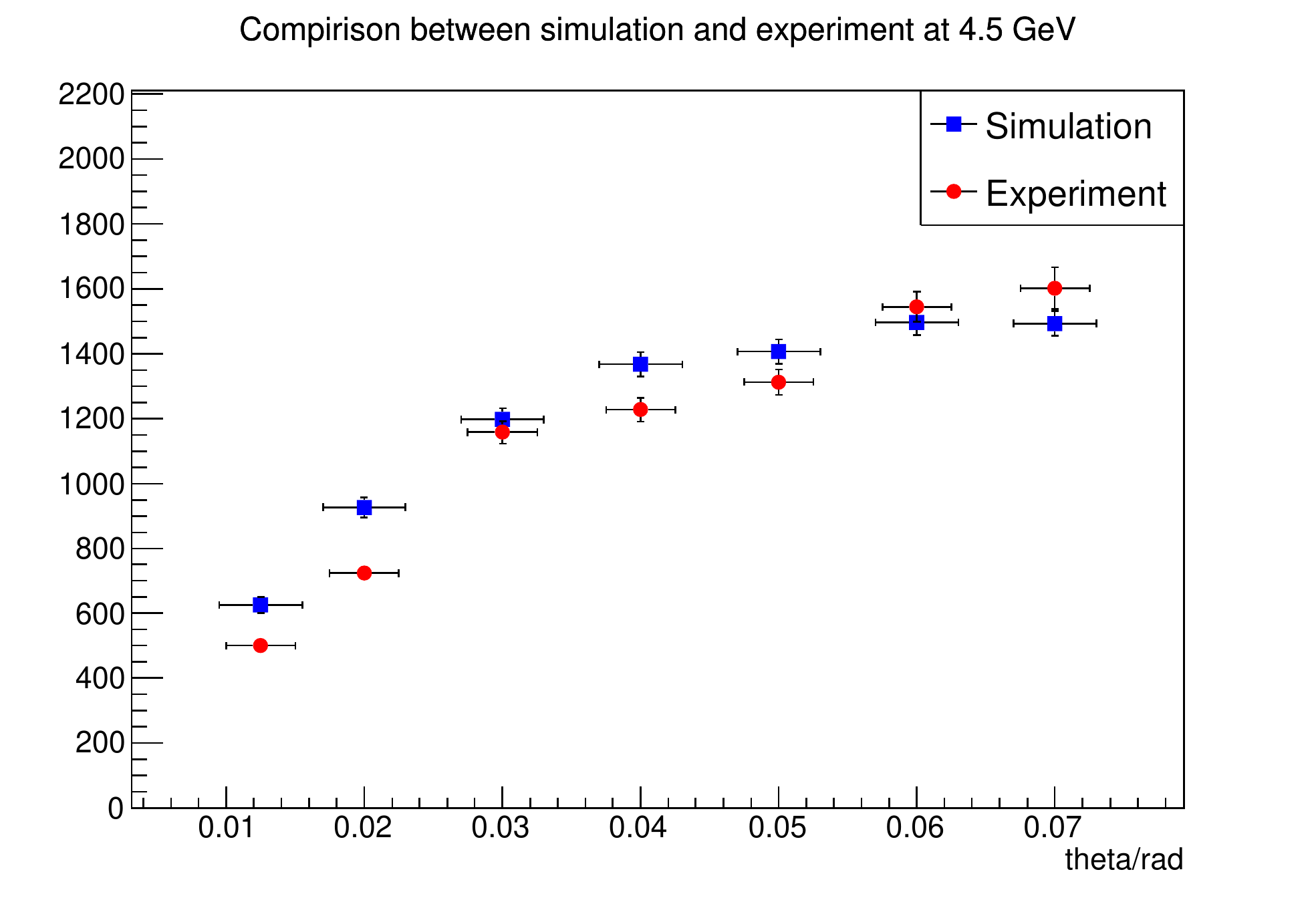}}
	\subfigure[]{ \includegraphics[width=0.45\textwidth]{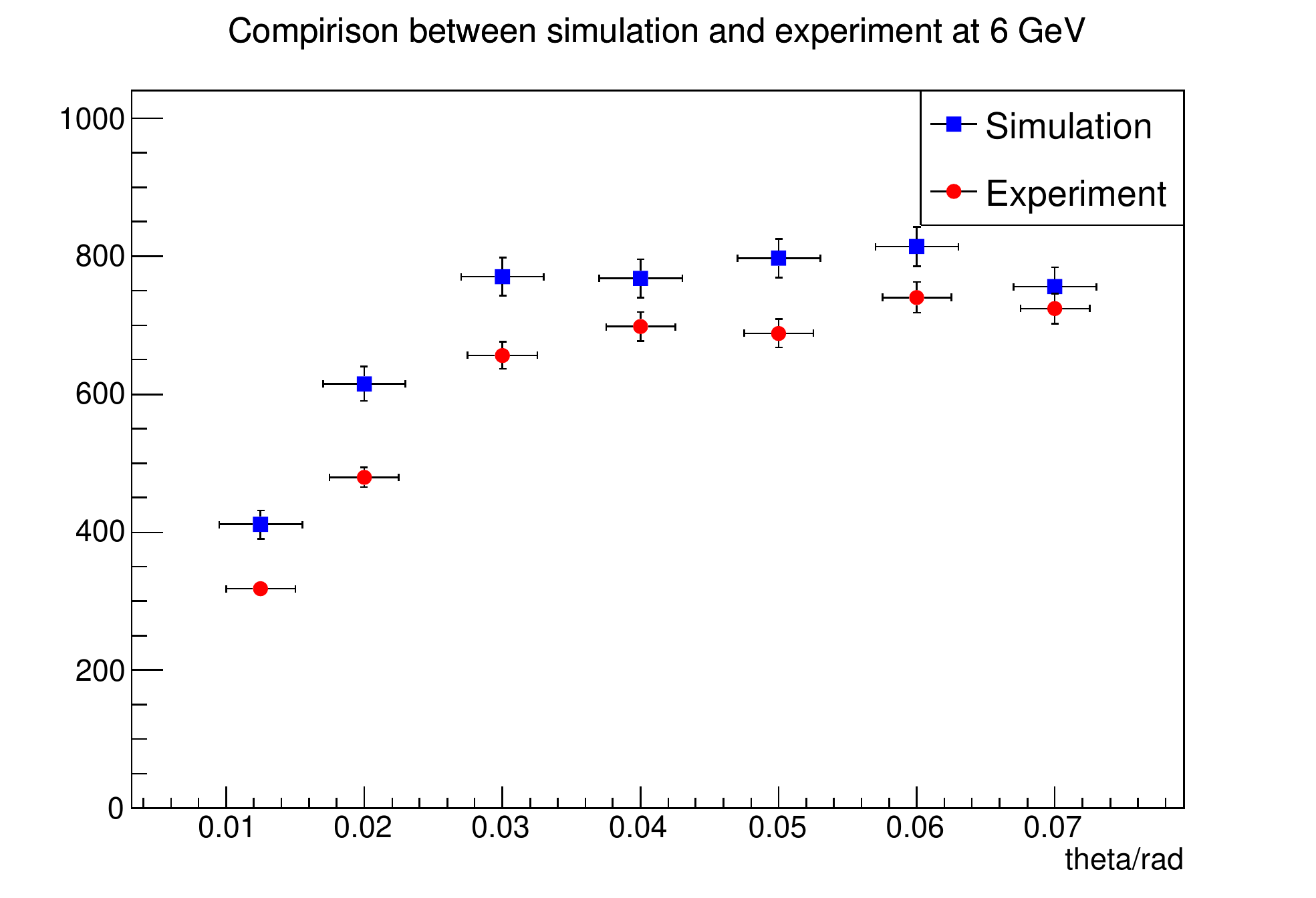}}
	\subfigure[]{ \includegraphics[width=0.45\textwidth]{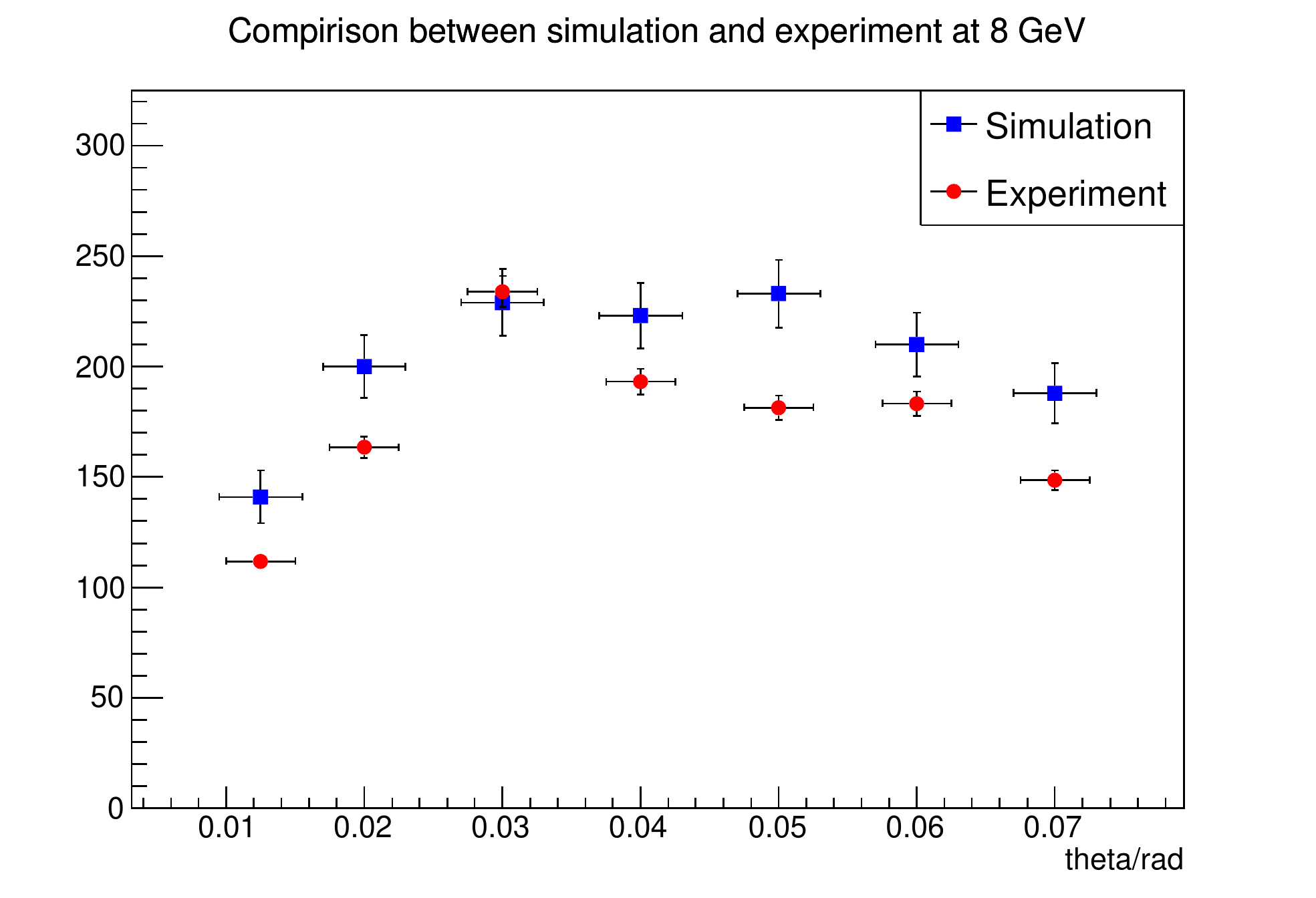}}
	\subfigure[]{ \includegraphics[width=0.45\textwidth]{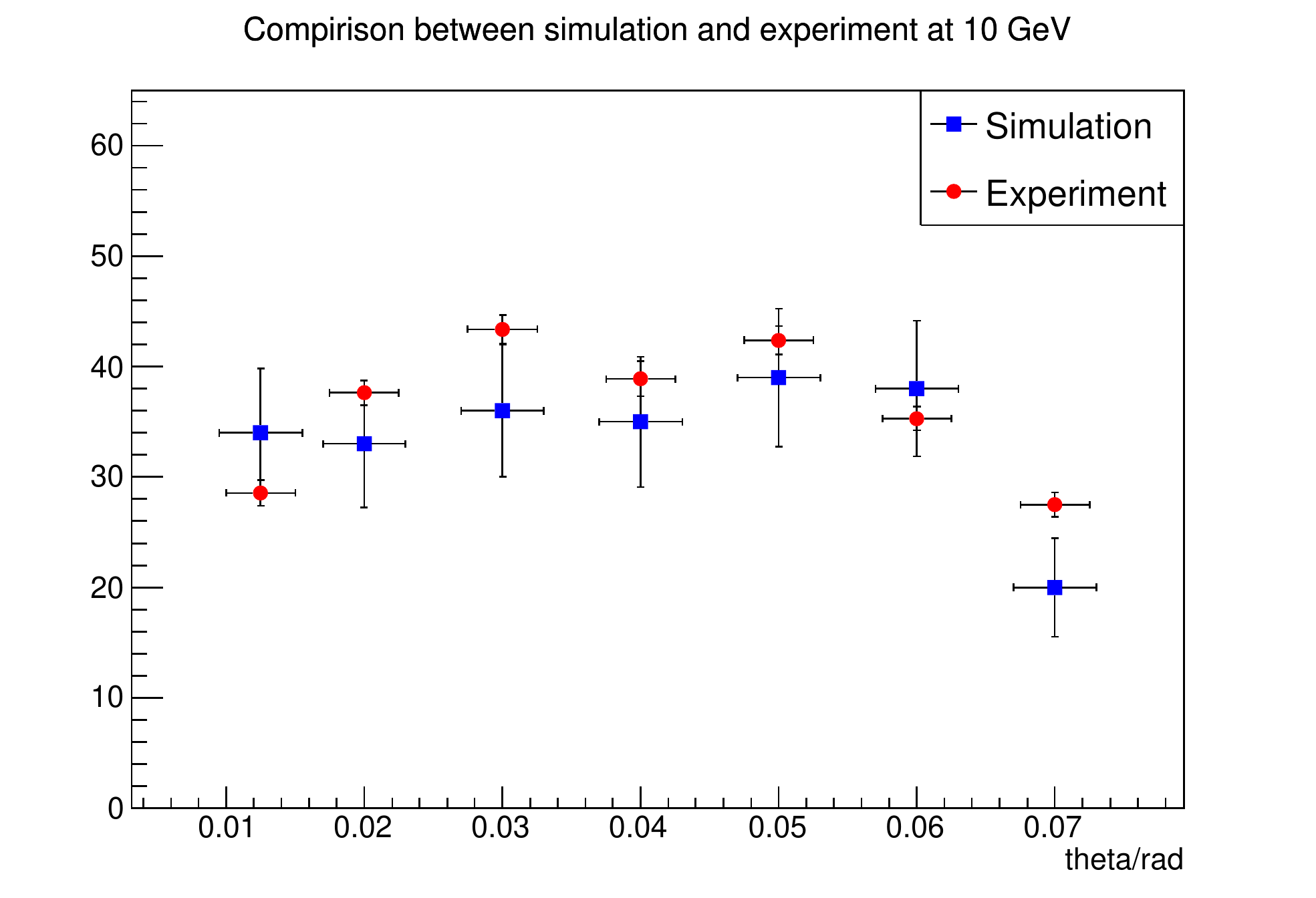}}
	\doublespacing
	\caption{The number of $\bar{p}$ simulated by this generator and the number calculated by measured cross sections when $\bar{p}$ escaping with (a) $4.5~{\rm{GeV}}/c$ (b) $6~{\rm{GeV}}/c$ (c) $8~{\rm{GeV}}/c$ (d) $10~{\rm{GeV}}/c$.}
	\label{fig:compare_CERN}
\end{figure}

In this experiment, a liquid hydrogen target, which is $10~\rm{cm}$ long and $2~\rm{cm}$ in radius, is collided by protons with $19.2~{\rm{GeV}}/c$ momenta. $\bar{p}$ are messured at laboratory angles of $12.5,~20,~30,~40,~50,~60,~70~\rm{mrad}$ with momenta of $4.5,~ 6,~8,~10,~11,~12,~13,~14~{\rm{GeV}}/c$. We build the same set-up as the experiment in simulation program and shoot $5\times10^8$ protons on the target. In order to collect more $\bar{p}$, the production cross section has been amplified by 2000 times. We use the same method as the experiment to get the yields in each angle and momentum of $\bar{p}$. Then both the expected yields of $\bar{p}$ calculated by experimental results and the yields of simulation are got and shown in Table~\ref{tab:compare_CERN}. To compare more intuitively, we draw them in the same picture Figure~\ref{fig:compare_CERN}. As we can see, the yields of simulation are consistent well with those calculated by experimental results.

\begin{table}[!htbp]
	\doublespacing
	\centering
	\begin{tabular}{cccc}
		\hline
		\hline
		$P({\rm GeV}/c)$ & $\theta$(mrad) & $N_{sim}$ & $N_{mea}$ \\
		\hline
		\multirow{7}*{4.5}  & 12.5	&	$625.00\pm25.00$	&	$500.62\pm15.02$ \\
				~  & 20.0	&	$926.00\pm30.43$	&	$723.76\pm21.71$ \\
				~  & 30.0	&	$1197.00\pm34.60$	&	$1157.92\pm34.74$ \\
				~  & 40.0	&	$1367.00\pm36.97$	&	$1227.01\pm36.81$ \\
				~  & 50.0	&	$1406.00\pm37.50$	&	$1312.16\pm39.36$ \\
				~  & 60.0	&	$1496.00\pm38.68$	&	$1543.48\pm46.30$ \\
				~  & 70.0	&	$1493.00\pm38.64$	&	$1601.07\pm64.04$ \\
		\hline
		\multirow{7}*{6.0}  & 12.5	&	$411.00\pm20.27$	&	$317.66\pm9.53$ \\
				~  & 20.0	&	$615.00\pm24.80$	&	$479.29\pm14.38$ \\
				~  & 30.0	&	$770.00\pm27.75$	&	$656.15\pm19.68$ \\
				~  & 40.0	&	$768.00\pm27.71$	&	$697.93\pm20.94$ \\
				~  & 50.0	&	$797.00\pm28.23$	&	$688.24\pm20.65$ \\
				~  & 60.0	&	$814.00\pm28.53$	&	$740.39\pm22.21$ \\
				~  & 70.0	&	$756.00\pm27.50$	&	$724.34\pm21.73$ \\
		\hline
		\multirow{7}*{8.0}  & 12.5	&	$141.00\pm11.87$	&	$111.78\pm3.35$ \\
				~  & 20.0	&	$200.00\pm14.14$	&	$163.41\pm4.90$ \\
				~  & 30.0	&	$229.00\pm15.13$	&	$234.00\pm7.02$ \\
				~  & 40.0	&	$223.00\pm14.93$	&	$193.06\pm5.79$ \\
				~  & 50.0	&	$233.00\pm15.26$	&	$181.39\pm5.44$ \\
				~  & 60.0	&	$210.00\pm14.49$	&	$183.29\pm5.50$ \\
				~  & 70.0	&	$188.00\pm13.71$	&	$148.53\pm4.46$ \\
		\hline
		\multirow{7}*{10.0}  & 12.5	&	$34.00\pm5.83$		&	$28.55\pm1.14$ \\
				~  & 20.0	&	$33.00\pm5.74$		&	$37.64\pm1.13$ \\
				~  & 30.0	&	$36.00\pm6.00$		&	$43.37\pm1.30$ \\
				~  & 40.0	&	$35.00\pm5.92$		&	$38.89\pm1.56$ \\
				~  & 50.0	&	$39.00\pm6.24$		&	$42.39\pm1.27$ \\
				~  & 60.0	&	$38.00\pm6.16$		&	$35.29\pm1.06$ \\
				~  & 70.0	&	$20.00\pm4.47$		&	$27.49\pm1.10$ \\
		\hline
	\end{tabular}
	\caption{Number of $\bar{p}$ recorded in simulation and the expected number calculated by cross sections measured in CERN ($5\times10^8$ incident proton with cross sections amplified by 2000 times). $P$ is the momenta of $\bar{p}$ and $\theta$ is the angle between the direction of escaping $\bar{p}$ and the incident direction of proton. $N_{sim}$ and $N_{mea}$ denote the number of events evaluated by simulation and that calculated by measured cross section respectively.}
	\label{tab:compare_CERN}
\end{table}

\subsection{Data taken by ITEP in 1994}

\begin{figure*}[!htbp]
	\doublespacing
	\centering
	\subfigure[]{ \includegraphics[width=0.45\textwidth]{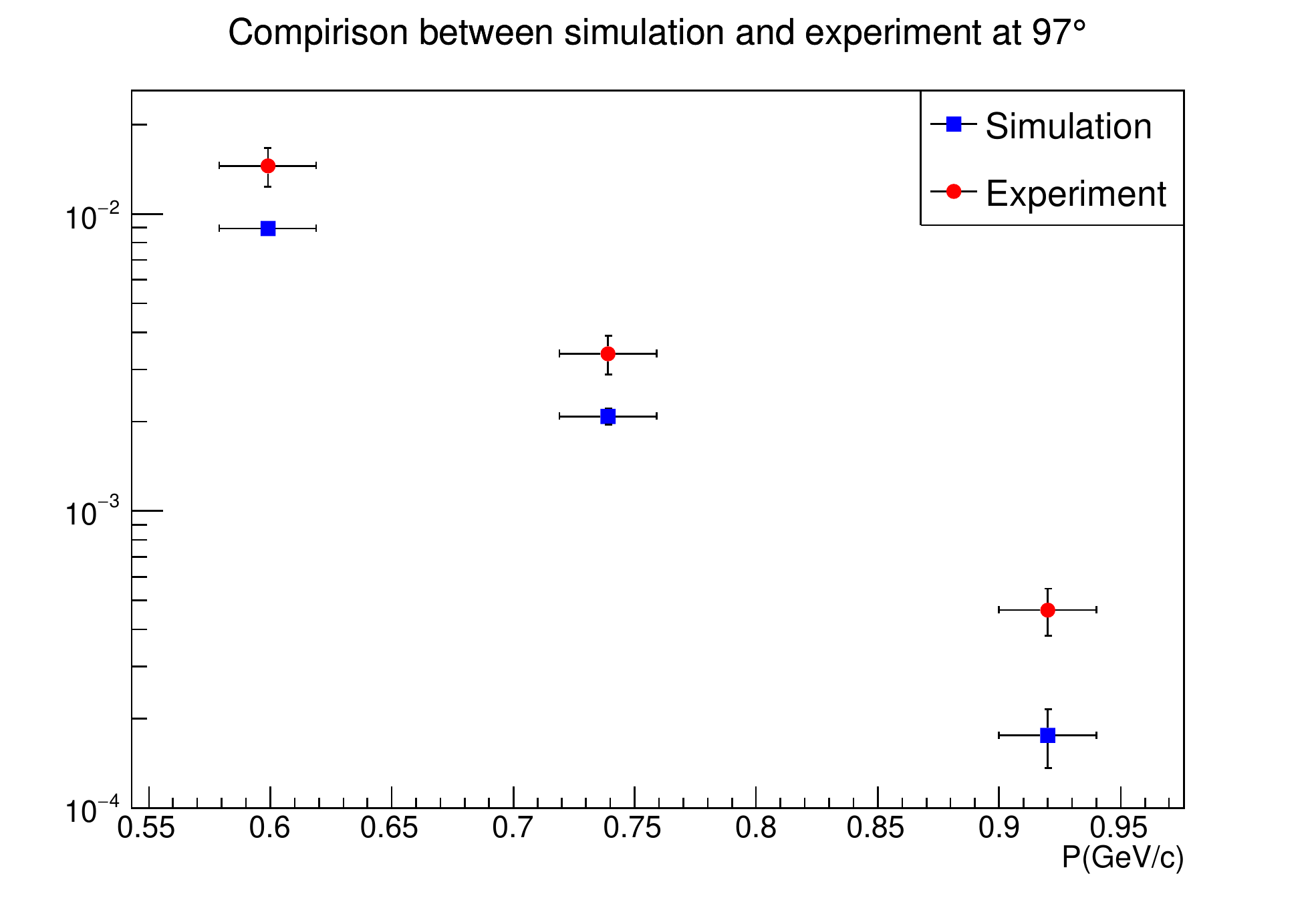}}
	\subfigure[]{ \includegraphics[width=0.45\textwidth]{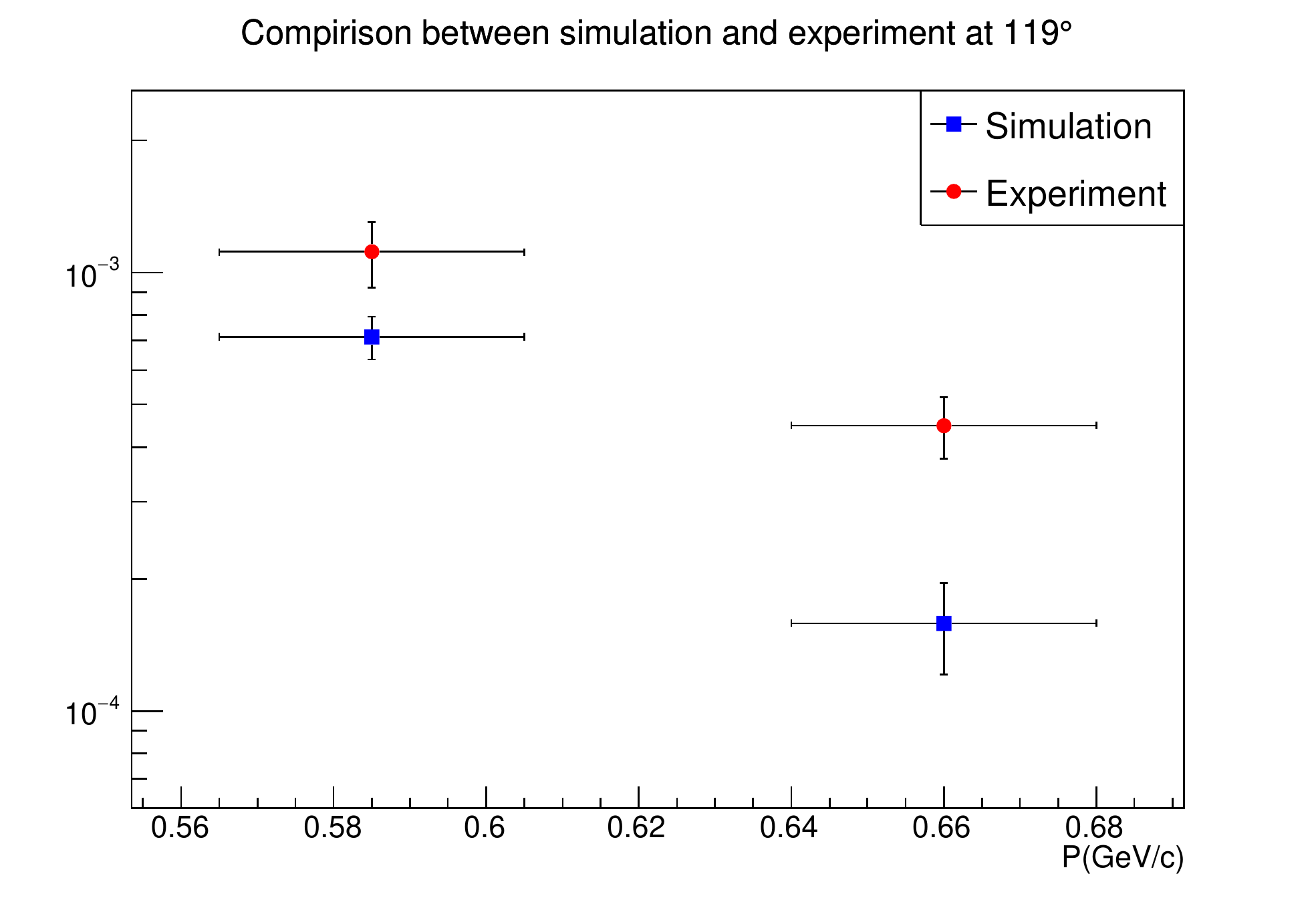}}
	\caption{The number of $\bar{p}$ simulated by this generator and the number calculated by measured cross sections when $\bar{p}$ escaping with (a) $97^{\circ}$ (b) $119^{\circ}$.}
	\label{fig:compare_ITEP}
\end{figure*}

Although the simulation results are very close to the measurements by CERN, the collection angles in CERN experiment are much smaller than the case of COMET, Mu2e and other experiment. Meanwhile, beam energy is much higher. In order to verify our generator with an experiment more similar with COMET, we use the ITEP results in 1994~\cite{Boyarinov:1994tp} as one more reference, where thin Be, Al, Cu and Ta foils are used as target hit by protons whose kinetic energy is $10.14~\rm{GeV}$. $\bar{p}$ are collected at laboratory angle $97^{\circ}$ and $119^{\circ}$. That is to say, the $\bar{p}$ flying backward have been measured in this experiments. In COMET experiment, $\bar{p}$ will be collected at much larger angle in laboratory system, which has never been measured or reported in history. However, the integral cross section of $\bar{p}$ drops sharply when the c.m.s energy decreases. Moreover, there are very few $\bar{p}$ scattered backward, which are about $1\times10^{-6}$ or less of those scattered forward. Therefore, the comparison is very difficult in this case. The cross section in simulation is amplified by $1\times10^7$ times. We shoot $5\times10^8$ protons to the target and get the total yield of $\bar{p}$. Meanwhile, a joint distribution of angle and momentum of escaping $\bar{p}$ in the ITEP case considering the momenta of nucleons is got by quick sampling for $5\times10^{12}$ times without simulation. After normalizing this joint distribution to the total yield of simulation, we can got an evaluated number of events in each point of ITEP experiment. At last the number got by simulation is compared with the expected yield calcualted by experimental cross sections. The comparison results have been shown in Table~\ref{tab:compare_ITEP}. Similar with CERN case, we draw each point at the same angle in the same picture, which can be seen in Figure~\ref{fig:compare_ITEP}. Simulation yields are consistent with experiment results within 56\%, which is acceptable for the background evaluation of COMET or Mu2e considering there will be no $\bar{p}$ produced if we use Geant4 official physics lists. However, the simulation yields are lower than the experiment results by $22\% \sim 56\%$ systematically, while the discrepancy will still be covered by the uncertainties within $3\sigma$, which will be further varified in COMET.

\begin{table}[!htbp]
	\doublespacing
	\centering
	\begin{tabular}{cccc}
		\hline
		\hline
		$\theta$(degree) & $P({\rm GeV}/c)$ & $N_{sim}$ & $N_{mea}$ \\
		\hline
		\multirow{3}*{$97^{\circ}$}  & 0.599       &       $(1.59\pm0.05)\times10^{-2}$        &       $(2.41\pm0.36)\times10^{-2}$ \\
					~  & 0.739       &       $(4.11\pm0.26)\times10^{-3}$        &       $(5.62\pm0.84)\times10^{-3}$ \\
					~  & 0.920       &       $(3.71\pm0.77)\times10^{-4}$        &       $(7.72\pm1.39)\times10^{-4}$ \\
		\hline
		\multirow{2}*{$119^{\circ}$} & 0.585	&	$(1.45\pm0.15)\times10^{-3}$		&	$(1.85\pm0.31)\times10^{-3}$ \\
					~  & 0.660	&	$(3.23\pm0.72)\times10^{-4}$		&	$(7.44\pm1.19)\times10^{-4}$ \\
		\hline
	\end{tabular}
	\caption{Number of $\bar{p}$ recorded in simulation and the expected number calculated by cross sections measured in ITEP ($5\times10^8$ incident proton with cross sections amplified by $1\times10^7$ times). $P$ is the momenta of $\bar{p}$ and $\theta$ is the angle between the direction of escaping $\bar{p}$ and the incident direction of proton. $N_{sim}$ and $N_{mea}$ denote the number of events evaluated by simulation and that calculated by measured cross section respectively.}
	\label{tab:compare_ITEP}
\end{table}

\subsection{Discussion about the discrepancy between experiments and the generator}

There are two main reasons for the discrepancy. Firstly, the model for the momentum distribution of nucleons used in our work is the Fermi Gas model~\cite{Benhar:1994hw}, which is also the common model in Geant4 simulation. However, Fermi Gas model may not be accurate. Secondly, the generator in this work is developed based on a theoretical model raised by Tan et al.~\cite{Tan:1983kgh} while this model may not be so accurate when $\bar{p}$ is produced at a large angle.

As is explained in our article, several theoretical models have been developed in history, but most of them are phenomenological models. That is to say, they don't purely start from standard model but are parameterized formulas modelling existing experimental results. However, the measurements of cross sections when $\bar{p}$ is produced at large angle and near-threhold energy are extramely lacked, which might results in the discrepancy between theory and experiment. Actually, there are multiple reasons for the difficulty in theoretical calculation:

\begin{itemize}
        \item[-] The models of nucleus may not be so accurate. Nucleus are very complicated systems including complex potential fields and interactions, which makes calculation difficult.
        \item[-] The process in our work $p+p \rightarrow p+p+p+\bar{p}$ is a hadronic process. When the $\sqrt{s}$ is much larger than threshold, the cross section can be calculated precisely by the perturbation theory. However, when we consider the production of $\bar{p}$ at a near-threshold energy, the non-pertuabative effects will significantly influent the calculation so that the cross sections cannot be directly calculated by standard model.
\end{itemize}

With more precise cross sections measured and more theoretical models raised in future, more accurate simulations will definitely be expected.

\section{Compare with Geant4 official process}

\begin{figure}[!htbp]
        \doublespacing
        \centering
        \includegraphics[width=0.7\textwidth]{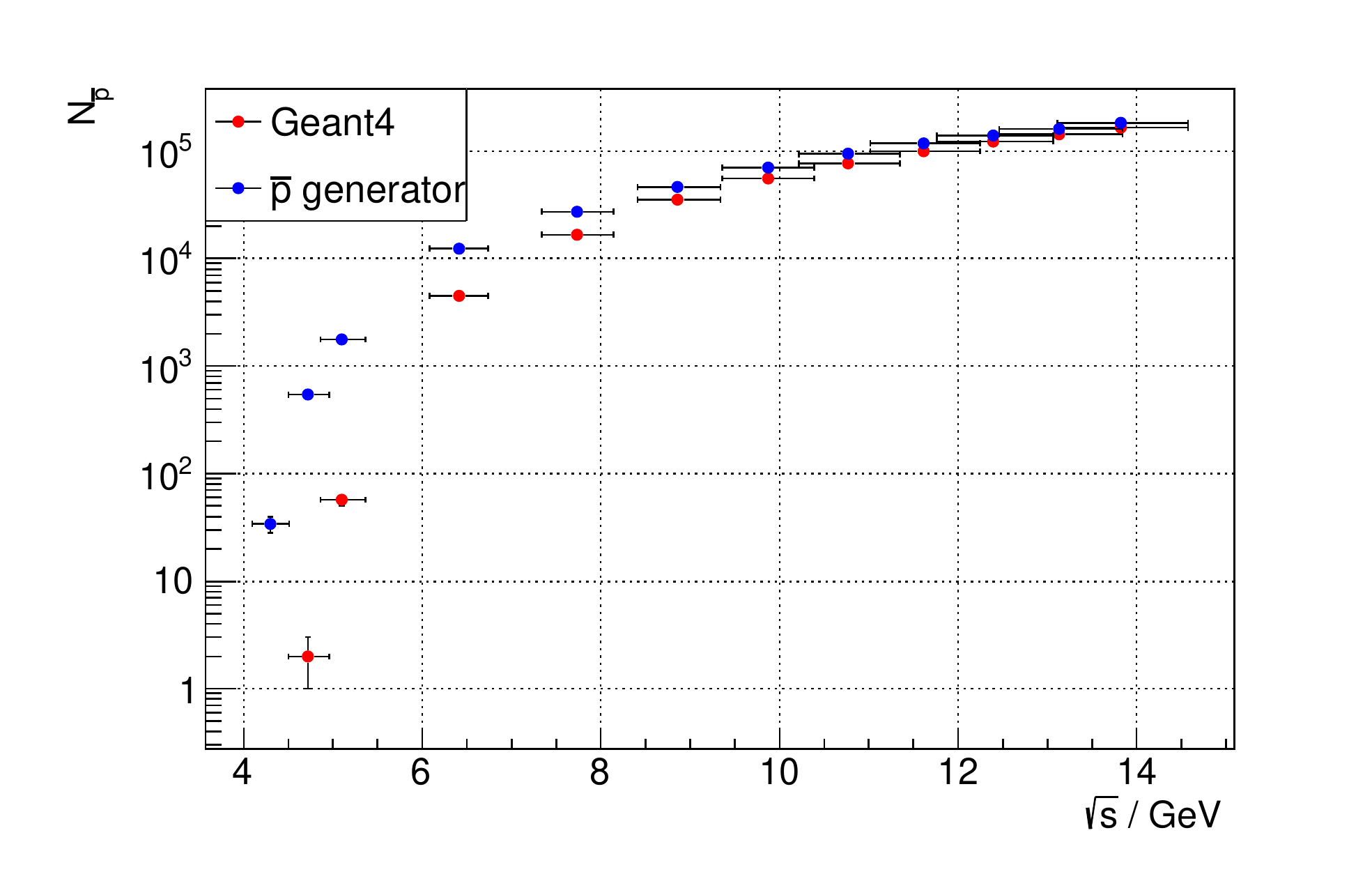}
        \caption{The number of events using Geant4 official physics list and the generator in this work when $1\times10^{7}$ incident protons hit the graphite target.}
        \label{fig:comparison_Geant4}
\end{figure}

Despite the discrepancy mentioned above, our generator significantly improves the accuracy of simulation comparing with the Geant4 official process. We shoot $1\times10^7$ protons with 8~GeV, 10~GeV, ..., 100~GeV kinetic energy to a graphite target (70~cm in length and 13~mm in radius) and simulate the $p$-nucleon collisions using Geant4 official physics list and our generator, respectively. Then the $\bar{p}$ produced by $p+p\rightarrow p+p+p+\bar{p}$ process are collected. Table~\ref{tab:comparison_Geant4} shows the simulation results. The yields of $\bar{p}$ of Geant4 official process are much lower than our generator when the $\sqrt{s}$ energy is close to the threshold and will gradually get close to our generator when energy gets higher, especially when $\sqrt{s}>10~{\rm GeV}$, which also varifies that the radial-scaling limit is already reached when $\sqrt{s}>10~{\rm GeV}$~\cite{Taylor:1975tm}. However, in the energy region of COMET~\cite{COMET:2009qeh, COMET:2018auw}, mu2e~\cite{Kutschke:2009zz} and other future experiments searching for CLFV, the yield of $\bar{p}$ simulated by Geant4 will be smaller than the yield of our generator by at least two magnitudes. To be clear, we draw the yields in each energy points into Figure~\ref{fig:comparison_Geant4}.

Considering that the yields of our generator when $\bar{p}$ escape at a small angle is accurate according to the comparison with CERN experiment and almost all the $\bar{p}$ will escape at a small angle due to the strong boost, the total yield of our generator will be much closer to the reality, which implies that the results by Geant4 official process is far from correct at the near-threshold energy. Therefore, our generator has significantly improved the accuracy of simulation of this process.

\begin{table}[!htbp]
        \doublespacing
        \centering
        \begin{tabular}{ccccc}
                \hline
                \hline
                $E_{k}^{proton}$ (GeV)  &       $\sqrt{s}$ (Gev)        &       $N_p$           &       $N_{\bar{p}}^{Geant4}$  &       $N_{\bar{p}}^{generator}$       \\
                \hline
                8                       &       $4.30_{-0.20}^{+0.21}$  &       $1\times10^7$   &       0                       &       $34\pm6$                        \\
                10                      &       $4.72_{-0.22}^{+0.24}$  &       $1\times10^7$   &       $2\pm1$                 &       $541\pm23$                      \\
                12                      &       $5.10_{-0.24}^{+0.26}$  &       $1\times10^7$   &       $57\pm7$                &       $1762\pm42$                     \\
                20                      &       $6.41_{-0.33}^{+0.33}$  &       $1\times10^7$   &       $4496\pm67$             &       $12408\pm111$                   \\
                30                      &       $7.73_{-0.39}^{+0.41}$  &       $1\times10^7$   &       $16649\pm129$           &       $27217\pm165$                   \\
                40                      &       $8.86_{-0.45}^{+0.48}$  &       $1\times10^7$   &       $35229\pm188$           &       $46276\pm215$                   \\
                50                      &       $9.87_{-0.51}^{+0.52}$  &       $1\times10^7$   &       $55496\pm236$           &       $70144\pm265$                   \\
                60                      &       $10.77_{-0.55}^{+0.58}$ &       $1\times10^7$   &       $76946\pm277$           &       $93908\pm306$                   \\
                70                      &       $11.61_{-0.59}^{+0.63}$ &       $1\times10^7$   &       $99539\pm315$           &       $117367\pm343$                  \\
                80                      &       $12.39_{-0.63}^{+0.67}$ &       $1\times10^7$   &       $121687\pm349$          &       $139588\pm374$                  \\
                90                      &       $13.13_{-0.67}^{+0.71}$ &       $1\times10^7$   &       $143663\pm379$          &       $161411\pm402$                  \\
                100                     &       $13.82_{-0.71}^{+0.75}$ &       $1\times10^7$   &       $164958\pm406$          &       $183864\pm429$                  \\
                \hline
                \hline
        \end{tabular}
        \caption{The number of events using Geant4 official physics list and the generator in this work when $1\times10^{7}$ incident protons hit the graphite target. $E_{k}^{proton}$ denotes the kinetic energy of incident protons. $\sqrt{s}$ is the center-of-mass energy whose uncertainty is evaluated by varying the momenum of nucleon within $[-0.1,0.1]~{\rm GeV}/c$. $N_p$ refers to the number of incident protons. $N_{\bar{p}}^{Geant4}$ and $N_{\bar{p}}^{generator}$ are the number of $\bar{p}$ produced by Geant4 official physics list and the generator in this work, respectively. Statistical uncertainty of $\bar{p}$ yield is considered in this table.}
        \label{tab:comparison_Geant4}
\end{table}

\section{Summary}

A generator of $\bar{p}$ produced in proton-nucleon interaction has been developed. By comparing with CERN and ITEP experiment results, the generator is in a fairly good agreement with the measurements when c.m.s energy is close to the threshold of $\bar{p}$ production. According to the comparison with experimental results, the yield when $\bar{p}$ is produced at small angle is accurate while the yield at much larger angle ($>90^{\circ}$) will have a discrepancy of $22\% \sim 56\%$. The discrepancy will be further tested by the COMET experiment in future. Using our generator, the production process of $\bar{p}$ in the collision of proton and nucleon can be simulated more accurate and, natually, background caused by $\bar{p}$ can be evaluated properly. In addition, as the parameterized formula can also describe the cross sections of other hadrons like $\pi$ and $K$ mesons, the method in this work may be used for more extensive simulations. Meanwhile, many more theoretical models have been raised in the past several decades~\cite{Danielewicz:1990cd, Sibirtsev:1997mq, Duperray:2003bd, diMauro:2014zea}, which may also be taken into consideration in our future work.

Besides, our generator is not limited for COMET experiment. It can be used in any other similar experiments that use Geant4 as their simulation toolkit in order to get the $\bar{p}$ yield near the threshold.

\section*{Acknowledgements}
The authors would like to express thanks to Yao Zhang and Ye Yuan for their strong support, significant effort and precious comments. The authors are grateful for members of COMET collaboration for their providing crucial information and valuable help during our work. 

This work is supported by the National Natural Science Foundation of China (NSFC) under Contracts Nos. 11935018, U1832204; International Partnership Program of Chinese Academy of Sciences, Grant No. 113111KYSB20190035





\bibliographystyle{elsarticle-num}
\bibliography{unsrt}

\end{document}